\documentclass[envcountsame,envcountsect]{llncs}

\RequirePackage[utf8]{inputenc}
\RequirePackage[T1]{fontenc}
\RequirePackage{lmodern}
\usepackage{xcolor}
\RequirePackage[english]{babel}

\RequirePackage{amsmath}
\RequirePackage{amsfonts}
\RequirePackage{amssymb}
\RequirePackage{mathrsfs}
\RequirePackage{stmaryrd}

\RequirePackage{xcolor}

\RequirePackage{multicol}

\RequirePackage{tikz}
  \usetikzlibrary{backgrounds}
  \usetikzlibrary{calc}
  \usetikzlibrary{fadings}
  \usetikzlibrary{positioning}
  \usetikzlibrary{decorations.pathreplacing}

\usepackage{calc}
\usepackage{url}
\usepackage{etoolbox}
\usepackage{options}
\RequirePackage{xparse}
\RequirePackage{xargs}
\RequirePackage{footnote}

\newcommand{\ToggleValue}[1]{%
  \iftoggle{#1}{true}{false}%
}

\colorlet{ColorM1}{black!80} 
\colorlet{ColorM2}{black!60} 

\colorlet{ColorS1}{black!70}
\colorlet{ColorS2}{black!50}
\colorlet{ColorS3}{black!30}

\colorlet{ColorB1}{black!12}
\colorlet{ColorB2}{black!80}

\usepackage[Language=English]{LanguageAB}
\usepackage{SpaceAB}
\usepackage{FrameAB}
\usepackage[Emph,In,Numbered]{AlgorithmAB}

\newcommand{\hex}[1]{\texttt{#1}}
\newcommand{\FF}{\mathbb{F}}

\newlength{\VSBox}
\newlength{\VXor}
\newlength{\VLin}
\newlength{\VBit}
\newlength{\VRound}
\newlength{\VXorPos}
\newlength{\VSBoxPos}
\newlength{\VLinPos}

\newlength{\VWord}
\newlength{\VWordPos}
\newlength{\VRCPos}
\newlength{\VBigBit}
\newlength{\VMerge}
\newlength{\VDiff}
\newlength{\VDiffPos}
\newlength{\HSBox}
\newlength{\HSBoxSep}
\newlength{\HXor}
\newlength{\HBlock}
\newlength{\HBitSep}

\newlength{\HFigSep}
\newlength{\HDiff}
\newlength{\HWordSep}
\newlength{\HWord}
\newlength{\Radius}
\newlength{\Tension}
\colorlet{CFill}{black!7}
\colorlet{CLine}{black!70}
\tikzset{
  BoxN/.style={
    draw=CLine,thick,rounded corners=\Radius,fill=CFill,text=black
  } ,
  LineN/.style={CLine} ,
  Merge1/.style={CLine,line width=1pt,>=stealth} ,
  Merge2/.style={CLine,line width=1.5pt,>=stealth} ,
  NameN/.style={black} ,
}
\newcommand{\BendTo}{.. controls +(0,-\Tension) and +(0,\Tension) ..}

\title{%
  Mathematical Backdoors in Symmetric Encryption Systems%
  \protect\footnote{This work has been presented at the FORmal Methods for Security Engineering (ForSE) conference 2017}  
  \subtitle{Proposal for a Backdoored AES-like Block Cipher}%
}
\titlerunning{Mathematical Backdoors in Symmetric Encryption Systems}
\author{Arnaud Bannier \and Eric Filiol}
\authorrunning{A. Bannier and E. Filiol}
\tocauthor{Arnaud Bannier and Eric Filiol}
\institute{%
  Operational Cryptology and Virology Lab, ESIEA,\\
  38 rue des Drs Calmette et Gu\'erin, 53000 Laval, France,\\
  \email{\{bannier, filiol\}@esiea.fr}%
}

\begin{document}
\maketitle


\begin{abstract}
Recent years have shown that more than ever governments and intelligence agencies try to control and bypass the cryptographic means used for the protection of data. Backdooring encryption algorithms is considered as the best way to enforce cryptographic control. Until now, only implementation backdoors (at the protocol/implementation/ma\-nagement level) are generally considered. In this paper we propose to address the most critical issue of backdoors: mathematical backdoors or by-design backdoors, which are put directly at the mathematical design of the encryption algorithm. While the algorithm may be totally public, proving that there is a backdoor, identifying it and exploiting it, may be an intractable problem. We intend to explain that it is probably possible to design and put such backdoors. Considering a particular family (among all the possible ones), we present BEA-1, a block cipher algorithm which is similar to the AES and which contains a mathematical backdoor enabling an operational and effective cryptanalysis. The BEA-1 algorithm (80-bit block size, 120-bit key, 11 rounds) is designed to resist to linear and differential cryptanalyses. A challenge will be proposed to the cryptography community soon. Its aim is to assess whether our backdoor is easily detectable and exploitable or not.
\end{abstract}

\section{Introduction}\label{sec:introduction}%\vspace{-1.5mm}

\noindent Despite the fact that in the late 90s/early 2000s, citizens have partially obtained the freedom for using cryptography, the recent years have shown that more than ever, governments and intelligence agencies still try to control and bypass the cryptographic means used for the protection of data and of private life. Snowden's leaks were a first upheaval. A tremendous number of secret projects (from NSA, GCHQ) have been revealed to the public opinion which proves this situation. 

While the need for the security of everyday life activities (for companies, for citizens) requires more and more cryptography, recent bothering initiatives by political decision-makers ask for an even stronger control over cryptography not to say preparing  the simple prohibition or ban of cryptographic application such as telegram. At the same time, the EU as well as a number of security agencies (such as French ANSSI, German BSI) confirmed that it was nonsense and militate for the mandatory use of end-to-end encryption.  

The recurring approaches and attempts consist in making the implementation of backdoors mandatory. The simplest and naive approach consists in enforcing key escrowing at the operators' level. But point-to-point encryption solutions (which are not equal to end-to-end encryption) like Telegram or Proton mail enable to prevent it. A number of different backdoor techniques are regularly mentioned or proposed. 

The most critical aspect in implementation backdoors lies on the fact that hackers or analysts may find them more or less easily and worse may exploit them. This is the reason why it is likely that IT operators or developers are very reluctant to accept backdoors until now. In case of leak, they will inevitably lose users' confidence and favor the development of trusted services abroad. In fact, the backdoor issue arises due to the fact that only implementation backdoors (at the protocol/implementation/management level) are generally considered.

In this paper we address the most critical issue of backdoors: mathematical or by-design backdoors. In other words, the backdoor is put directly at the mathematical design of the encryption algorithm. While the algorithm may be totally public, proving that there is a backdoor, identifying it and exploiting it, may be an intractable problem, unless you know the backdoor. To some extent, the RSA's \texttt{Dual\_EC\_DRBG} standard case falls within this category \cite{shumow2007possibility}. Other non-public examples are known within the military cryptanalysis community, and partially revealed to the public thanks to the 1995 Hans Buehler case \cite{strehle1994verschlusselt}. This kind of backdoor is the most difficult one to address and there is quite no public work on that topic. It is generally the technical realm of a few among the most eminent intelligence agencies (namely NSA, GCHQ, SVR/GRU) which moreover have the ability and power to step in and to influence the international standardization processes.

We intend to explain that it is probably possible to design and put such backdoors. Considering a particular case of mathematical backdoors (among all the possible ones) based on our previous work \cite{cryptoeprint:2016:493}, we present a block cipher algorithm which is similar to the AES and which contains a mathematical backdoor enabling an operational and effective cryptanalysis (in other words in a limited time on a modern desktop computer and with a limited number of plaintext/ciphertext pairs). This block cipher algorithm (80-bit block, 120-bit key size, 11 rounds) is designed to resist to linear and differential cryptanalyses.

This paper is organized as follows. In Section~\ref{s1} we explore the concept of backdoors and trapdoors and we identify two main categories, each containing itself subcategories depending on the nature of the cipher (stream or block ciphers). This observation is backed by the personal experience of the second author as a military cryptanalyst. We also present the state-of-the-art, history and previous work regarding backdoors, mostly in symmetric cryptography. In Section~\ref{s2}, we present our backdoored block cipher algorithm BEA-1 (standing for \textit{Backdoored Encryption Algorithm 1}), based on our work \cite{cryptoeprint:2016:493}. This is a particular family of trapdoors using a suitable partition of the plaintext and ciphertext spaces. In Section~\ref{s3}, we discuss the cryptographic security of this cipher, with respect to linear and differential cryptanalyses. We also propose a cryptographic challenge to the cryptography community%\footnote{This challenge will be launched on the Arxiv repository right after the conference, at the end of February.}
, regarding the backdoor identification and exploitation. We suppose that this backdoor is likely to be detected. Such a challenge should enable to prove or disprove this claim. Lastly we conclude in Section~\ref{s4} and explore future work.

\section{The Concept of Backdoor} \label{s1}%\vspace{-1.5mm}
\subsection{Definition and Classification Proposal}%\vspace{-1.5mm}
\noindent Trapdoors are a two-face, key concept in modern cryptography. It is primarily related to the concept of ``\textit{trapdoor function}'' \textemdash{} a function that is easy to compute in one direction, yet difficult to compute in the opposite direction without special information, called the ``\textit{trapdoor}''. This first ``face'' relates most of the time to asymmetric cryptography algorithms. It is a necessary condition to get reversibility between the sender/receiver (encryption) or the signer/verifier (digital signature). The trapdoor mechanism is always fully public and detailed. The security and the core principle is based on the existence of a secret information (the private key) which is essentially part of the trapdoor. In other words, the private key can be seen as \emph{the} trapdoor.

The second ``face'' of the concept of trapdoor relates to the more subtle and perverse concept of ``mathematical backdoor'' and is a key issue in symmetric cryptography (even if the issue of backdoors may be extended to asymmetric cryptography; see for example the case of the \texttt{DUAL EC\_DRBG} \cite{shumow2007possibility}, or the case of trapdoored primes addresses recently in \cite{cryptoeprint:2016:961}). 

In this case, the aim is to insert hidden mathematical weaknesses which enable one who knows them to break the cipher. If possible, these weaknesses should be independent of the secret key. Somehow, it consists to create a hidden asymmetry to the detriment of the legitimate users of the communication and to the benefit of the eavesdropper. In this context, the existence of a backdoor is a strongly undesirable property. 

In the rest of the present section, we will oppose the term of trapdoor (desirable property) to that of backdoor (undesirable property). While the term of trapdoor has been already used in the very few literature covering this second face of this problem, we suggest however to use the term of backdoor to describe the issue of hidden mathematical  weaknesses. This would avoid ambiguity and maybe would favor the research work around a topic which is nowadays mostly addressed by governmental entities in the context of cryptography control and regulations.

Inserting backdoors in encryption algorithms underlies quite systematically the choice of cryptographic standards (DES, AES\dots). The reason is that the testing, validation and selection process is always conducted by governmental entities (NIST or equivalent) with the technical support of secret entities (NSA or equivalent). So an interesting and critical research area is: ``how easy and feasible is it to design and to insert backdoors (at the mathematical level) in encryption algorithms?''. In this paper, we intend to address one very particular case of this question. It is important to keep in mind that a backdoor may be itself defined in the following two ways.
\begin{itemize}
\item As a ``natural weakness'' known \textemdash{} but non disclosed \textemdash{} only by the tester/validator/final decision-maker (e.g. the NSA as it could have been the case for the AES challenge). The best historic example is that 
of the differential cryptanalysis. Following Biham and Shamir's seminal work in 1991 \cite{biham1993differential}, NSA acknowledged that it was aware of that cryptanalysis years ago \cite{Schneier2013}. Most of experts estimate that it was nearly 20 years ahead. However a number of non public, commercial block ciphers in the early 90s may be weak with respect to differential cryptanalysis.
\item As an intended design weakness put by the author of the algorithm. To the authors knowledge, there is no known cases for public algorithms yet.
\end{itemize}

As far as symmetric cryptography is concerned, there are two major families of cipher systems for which the issue of backdoor must be considered differently.
\begin{itemize}
\item \textit{Stream ciphers}. Their design complexity is rather low since they mostly rely on algebraic primitives (LFSRs and Boolean functions which have intensely been studied in the open literature). Until the late 70s, backdoors relied on the fact that quite all algorithms were proprietary and hence secret. It was then easy to hide non primitive polynomials, weak combining Boolean functions. The Hans Buehler case in 1995 \cite{strehle1994verschlusselt} shed light on that particular case.
\item \textit{Block ciphers}. This class of encryption algorithms is rather recent (end of the 70s for the public part). They exhibit so a huge combinatorial complexity that it is reasonable to think to backdoors. As described in \cite{daemen2002design} for a $k$-bit secret key and a $m$-bit input/output block cipher there are $((2^m)!)^{2^k}$ possible such block ciphers. For such an algorithm, the number of possible internal states is so huge that we are condemned to have only a local view of the system, that is, the round function or the basic cryptographic primitives. We cannot be sure that there is no degeneration effect at a higher level. This  point has been addressed in \cite{daemen2002design} when considering linear cryptanalysis. Therefore, it seems reasonable to think that this combinatorial richness of block cipher may be used to hide backdoors.    
\end{itemize}
Since block ciphers are the most widely used encryption algorithms nowadays by the general public and the industry, we will focus on them in the rest of the paper. Backdoors in stream ciphers have quite never been exposed to the public.
%-----------------------------------------------------
\subsection{Previous Work}
One of the first trapdoor cipher was proposed in 1997 by Rijmen and Preneel \cite{rijmen1997family}. The S-boxes are selected randomly and then modified to be weak to the linear cryptanalysis. They are finally applied to a Feistel cipher such as CAST or LOKI91. But because of the big size of the S-boxes, the linear table of such an S-box cannot be computed. However the knowledge of the trapdoor gives a good linear approximation of the S-boxes which is then used in a linear cryptanalysis. As an example, the authors created a 64-bit block cipher based on CAST cipher, and four $8\times32$ S-boxes. If the parameters of the trapdoors are known, a probabilistic algorithm allows to recover the key easily. Such a family of trapdoor ciphers leads to recover only a part of the key, and the authors claim that the trapdoor is undetectable. But in \cite{wu1998cryptanalysis}, Wu and al. discovered a way to recover the trapdoor if the attacker knows its global design but not the parameters. They also showed that there exists no parameter allowing to hide the trapdoor. Nevertheless, it is worthwhile to mention that in practice, if a real cipher containing a trapdoor is given, the presence of the trapdoor will certainly not be revealed. 

In \cite{paterson1999imprimitive}, a DES-like trapdoor cipher exploiting a weakness induced by the round functions is presented. The group generated by the round functions acts imprimitively on the message space to allow the design of the trapdoor. In other words, this group preserves a partition of the message space between input and output of the round function. Such a construction leads to the design of a trapdoor cipher composed of 32 rounds and using a 80 bits key. The knowledge of the trapdoor allows to recover the key using $2^{41}$ operations and $2^{32}$ plaintexts. Even if the mathematical material to build the trapdoor is given, no general algorithm is detailed to construct such S-boxes. Furthermore, as the author says, S-boxes using these principles are incomplete (half of the cipher text bits are independent of half of the plaintext bits). Finally, the security against the differential attack is said \emph{not as high as one might expect}. 

More recently in \cite{angelova2013plaintext}, the authors created non-surjective S-boxes embedding a parity check to create a trapdoor cipher. The message space is thus divided into cosets and leads to create an attack on this DES-like cipher in less than $2^{23}$ operations. The security of the whole algorithm, particularly against linear and differential cryptanalyses is not given and the authors admit that their attack is dependent on the first and last permutation of the cipher. Finally, the non-surjective S-boxes may lead to detect easily the trapdoor by simply calculating the image of each input vector. This problem is naturally avoided in a Substitution-Permutation Network (SPN) in which S-boxes are bijective by definition.

In a slightly different context, Caranti and al. answer to Patterson's question by the affirmative in \cite{caranti2006imprimitive}, by proving that the imprimitivity of the group generated by round functions is actually related to the cosets of a linear subspace. They also give some conditions to create such a primitive group to design a secure cipher that cannot contain such trapdoor, and finally show that the AES respects these conditions. They add in \cite{caranti2009some} an algorithm to verify this last condition simply and show that AES and Serpent S-boxes verify this property.

%---------------------------------------------------------
\section{Description of BEA-1} \label{s2}
The algorithm BEA-1 (standing for \textit{Backdoored Encryption Algorithm version 1}) is based on our research work on partition-based trapdoors \cite{cryptoeprint:2016:493}.
This section is intended to describe this cipher precisely. The cipher operates on 80-bit data blocks using a 120-bit master key. Our algorithm is directly inspired by Rijndael \cite{daemen2002design}, the block cipher designed by Joan Daemen and Vincent Rijmen which is now known as the AES (the encryption standard proposed by the USA, under the hauspices of NIST and NSA \cite{nist_aes}). Consequently, our cipher is a Substitution-Permutation Network.

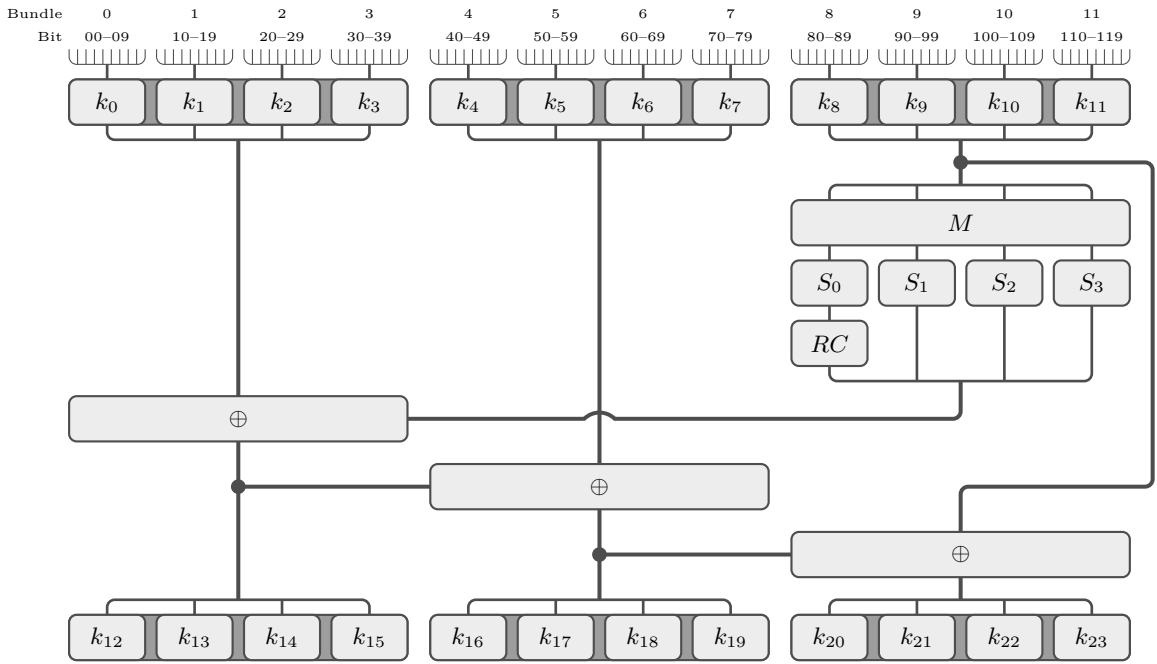
\begin{figure*}
  \centering
  \deflength{\VXor}      {6 mm}
  \deflength{\VSBox}     {\VXor}
  \deflength{\VDiff}     {\VXor}
  \deflength{\VWord}     {\VXor}
  \deflength{\VBit}      {2 mm}
  \deflength{\VBigBit}   {3 mm}
  \deflength{\VMerge}    {\VBit}
  \deflength{\VXorPos}   {\VBit}
  \deflength{\VDiffPos}  {\VWord +\VBit +2\VMerge +2\VBigBit}
  \deflength{\VSBoxPos}  {\VDiffPos +\VDiff +\VBit}
  \deflength{\VRCPos}    {\VSBoxPos +\VSBox +\VBit}
  \deflength{\VXorPos}   {\VRCPos +\VSBox +\VMerge +\VBit}
  \deflength{\VWordPos}  {\VXorPos +3\VXor +3\VBigBit +\VMerge}
  \deflength{\HSBox}     {10mm}
  \deflength{\HSBoxSep}  {1.5mm}
  \deflength{\HWordSep}  {3mm}
  \deflength{\HBitSep}   {0.111\HSBox}
  \deflength{\HXor}      {4\HSBox +3\HSBoxSep}
  \deflength{\HDiff}     {4\HSBox +3\HSBoxSep}
  \deflength{\HBlock}    {\HSBox +\HSBoxSep}
  \deflength{\HWord}     {4\HSBox +3\HSBoxSep}
  \deflength{\Radius}    {1 mm}
  \begin{tikzpicture}
  \useasboundingbox (0,8mm) rectangle (3\HWord +2\HWordSep,-\VWordPos -\VWord -2mm);
  % .................................................................................................
  % Coordonnées
  %
  \coordinate (Diff) at (2\HWord +2\HWordSep, -\VDiffPos);
  \coordinate (RC)   at (2\HWord +2\HWordSep, -\VRCPos);
  \foreach \X in {0,1,2}{
    \coordinate (WordA-\X) at (\X*\HWord +\X*\HWordSep, -\VBit);
    \coordinate (WordB-\X) at (\X*\HWord +\X*\HWordSep, -\VWordPos);
    \coordinate (Xor-\X) at (\X*\HWord +\X*\HWordSep, -\VXorPos -\X*\VXor - \X*\VBigBit);
    \coordinate (BigLineA-\X) at ($(WordA-\X)+(.5\HWord,-\VWord -\VMerge)$);
    \coordinate (BigLineD-\X) at ($(Xor-\X)+(.5\HXor,0)$);
    \coordinate (BigLineE-\X) at ($(Xor-\X)+(.5\HXor,-\VXor)$);
    \coordinate (BigLineF-\X) at ($(WordB-\X)+(.5\HWord,\VMerge)$);
  }
  \coordinate (BigLineB-2) at ($(Diff)+(.5\HDiff,\VMerge)$);
  \coordinate (BigLineC-2) at ($(RC)+(.5\HDiff,-\VSBox -\VMerge)$);
  \foreach \Y in {0,...,3}{
    \coordinate (BundleC-2\Y) at (2\HWord +2\HWordSep +.5\HSBox +\Y*\HBlock, -\VDiffPos);
    \coordinate (BundleD-2\Y) at (2\HWord +2\HWordSep +.5\HSBox +\Y*\HBlock, -\VDiffPos -\VDiff);
    \coordinate (BundleE-2\Y) at (2\HWord +2\HWordSep +.5\HSBox +\Y*\HBlock, -\VSBoxPos -\VSBox);
    \foreach \X in {0,1,2}{
      \coordinate (BundleA-\X\Y) at ($(WordA-\X)+(.5\HSBox +\Y*\HBlock,0)$);
      \coordinate (BundleB-\X\Y) at ($(WordA-\X)+(.5\HSBox +\Y*\HBlock,-\VWord)$);
      \coordinate (BundleF-\X\Y) at ($(WordB-\X)+(.5\HSBox +\Y*\HBlock,0)$);
    }
    \coordinate (SBox-\Y) at (2\HWord +2\HWordSep +\Y*\HBlock, -\VSBoxPos);
  }
  % .................................................................................................
  % Lignes et boites
  %
  \foreach \X in {0,1,2}{
    \draw[BoxN,fill=CLine!50!CFill] (WordA-\X) rectangle ++(\HWord,-\VWord);
    \draw[BoxN,fill=CLine!50!CFill] (WordB-\X) rectangle ++(\HWord,-\VWord);
    \foreach \Y in {0,...,3}{
      \draw[BoxN] ($(WordA-\X)+(\Y*\HBlock,0)$) rectangle ++(\HSBox,-\VSBox);
      \draw[BoxN] ($(WordB-\X)+(\Y*\HBlock,0)$) rectangle ++(\HSBox,-\VSBox);
      \draw[Merge1] (BundleA-\X\Y) -- ++(0,\VBit);
      \draw[LineN,rounded corners=\Radius] (\X*\HWord +\X*\HWordSep +\Y*\HBlock,\VMerge)
      -- ++(0,-\VMerge) -- ++(\HSBox, 0) -- ++(0,\VMerge);
      \foreach \Z in {1,...,8}{
        \draw[LineN] (\X*\HWord +\X*\HWordSep +\Y*\HBlock +\Z*\HBitSep,0) -- ++(0,\VMerge);
      }
    }
    \draw[LineN,rounded corners=\Radius,line width=1pt]
    (BundleB-\X0) -- ++(0,-\VMerge) -- ($(BundleB-\X3)+(0,-\VMerge)$) -- ++(0,\VMerge);
    \draw[Merge1]  (BundleB-\X1) -- ++(0,-\VMerge);
    \draw[Merge1]  (BundleB-\X2) -- ++(0,-\VMerge);
    \draw[BoxN] (Xor-\X) rectangle ++(\HDiff,-\VDiff);
    \draw[LineN,rounded corners=\Radius,line width=1pt]
    (BundleF-\X0) -- ++(0,\VMerge) -- ($(BundleF-\X3)+(0,\VMerge)$) -- ++(0,-\VMerge);
    \draw[LineN,line width=1pt]  (BundleF-\X1) -- ++(0,\VMerge);
    \draw[LineN,line width=1pt]  (BundleF-\X2) -- ++(0,\VMerge);
  }
  \foreach \Y in {0,...,3}{
    \draw[BoxN] (SBox-\Y) rectangle ++(\HSBox,-\VSBox);
  }
  \draw[LineN,rounded corners=\Radius,line width=1pt]
  (BundleC-20) -- ++(0,\VMerge) -- ($(BundleC-23)+(0,\VMerge)$) -- ++(0,-\VMerge);
  \draw[Merge1]  (BundleC-21) -- ++(0,\VMerge);
  \draw[Merge1]  (BundleC-22) -- ++(0,\VMerge);
  \draw[Merge1]  (BundleD-20) -- ++(0,-\VBit);
  \draw[Merge1]  (BundleD-21) -- ++(0,-\VBit);
  \draw[Merge1]  (BundleD-22) -- ++(0,-\VBit);
  \draw[Merge1]  (BundleD-23) -- ++(0,-\VBit);
  \draw[Merge1]  (BundleE-20) -- ++(0,-\VMerge);
  \draw[Merge1,rounded corners=\Radius]
  ($(BundleE-20)+(0,-\VBit -\VSBox)$) -- ++(0,-\VMerge) -- ++(3\HBlock,0) -- (BundleE-23);
  \draw[Merge1]  (BundleE-21) -- ++(0,-\VBit -\VSBox -\VMerge);
  \draw[Merge1]  (BundleE-22) -- ++(0,-\VBit -\VSBox -\VMerge);
  \draw[BoxN] (Diff) rectangle ++(\HDiff,-\VDiff);
  \draw[BoxN] (RC) rectangle ++(\HSBox,-\VSBox);
  \foreach \X in {0,1}{
    \draw[Merge2] (BigLineA-\X) -- (BigLineD-\X);
    \draw[Merge2] (BigLineE-\X) -- (BigLineF-\X);
    \draw[Merge2,fill] ($(BigLineE-\X)+(0,-\VBigBit -.5\VXor)$) circle(2pt);
    \draw[Merge2]      ($(BigLineE-\X)+(0,-\VBigBit -.5\VXor)$) -- ++(.5\HXor +\HWordSep, 0);
  }
  \draw[Merge2] (BigLineA-2) -- coordinate[midway] (tmp) (BigLineB-2);
  \draw[Merge2,fill] (tmp) circle(2pt);
  \draw[Merge2]{[rounded corners=\Radius]
    (BigLineC-2) |- ($(BigLineD-1)+(3mm,\VBigBit +.5\VXor)$)}
  -- ($(BigLineD-1)+(2mm,\VBigBit +.5\VXor)$) to[bend right=45] ++(-4mm,0)
  -- ($(Xor-0)+(\HXor,-.5\VXor)$);
  \draw[Merge2,rounded corners=\Radius]
  (tmp) -| ($(BigLineD-2)+(.5\HXor +3mm, \VBigBit +.5\VXor)$) -| (BigLineD-2);
  \draw[Merge2] (BigLineE-2) -- (BigLineF-2);
  %
  % .................................................................................................
  % Textes
  %
  \foreach \X/\Y/\TxtA/\TxtB in {%
    0/0/0/12, 0/1/1/13, 0/2/ 2/14, 0/3/ 3/15,%
    1/0/4/16, 1/1/5/17, 1/2/ 6/18, 1/3/ 7/19,%
    2/0/8/20, 2/1/9/21, 2/2/10/22, 2/3/11/23%
  }{
  \draw[NameN] ($(WordA-\X)+(\Y*\HBlock +.5\HSBox, -.5\VSBox)$) node {$k_{\TxtA}$};
  \draw[NameN] ($(WordB-\X)+(\Y*\HBlock +.5\HSBox, -.5\VSBox)$) node {$k_{\TxtB}$};
}
\foreach \X in {0,1,2}{
  \draw[NameN] ($(Xor-\X)+(.5\HXor,-.5\VXor)$) node {$\oplus$};
}
\foreach \Y in {0,1,2,3}{
  \draw[NameN] ($(SBox-\Y)+(.5\HSBox,-.5\VSBox)$) node {$S_{\Y}$};
}
\draw[NameN] ($(Diff)+(.5\HDiff,-.5\VDiff)$) node {$M$};
\draw[NameN] ($(RC)+(.5\HSBox,-.5\VSBox)$) node {$RC$};
\draw (0,0) node[above=2mm] {\tiny\makebox[0pt][r]{Bit~}};
\draw (0,0) node[above=5mm] {\tiny\makebox[0pt][r]{Bundle~}};
\foreach \X/\Y/\Txt in {%
  0/0/0, 0/1/1, 0/2/2, 0/3/3,%
  1/0/4, 1/1/5, 1/2/6, 1/3/7,%
  2/0/8, 2/1/9, 2/2/10, 2/3/11%
}{
\draw ($(BundleA-\X\Y)+(0,\VBit)$) node[above=2mm] {\tiny\Txt 0--\Txt 9};
\draw ($(BundleA-\X\Y)+(0,\VBit)$) node[above=5mm] {\tiny\Txt};
}
\end{tikzpicture}
\caption{A diagrammatic representation of the key schedule \texttt{ExpandKey}}\label{Fig:ExpKey}
\end{figure*}

The encryption consists in applying eleven times a simple keyed operation called \emph{round function} to the data block. A different 80-bit round key is used for each iteration of the round function. Since the last round is slightly different and uses two round keys, the encryption requires twelve 80-bit round keys.
These round keys are derived from the 120-bit master key using an algorithm called \emph{key schedule} (depicted in Figure~\ref{Fig:ExpKey}).

\begin{figure*}
  \centering
  \deflength{\VXor}      {6 mm}
  \deflength{\VSBox}     {\VXor}
  \deflength{\VDiff}     {\VXor}
  \deflength{\VLin}      {9 mm}
  \deflength{\VBit}      {2 mm}
  \deflength{\VXorPos}   {\VBit}
  \deflength{\VSBoxPos}  {\VXorPos +\VXor +\VBit}
  \deflength{\VLinPos}   {\VSBoxPos +\VSBox +\VBit}
  \deflength{\VDiffPos}  {\VXor +\VSBox +\VLin +4\VBit}
  \deflength{\VRound}    {\VXor +\VSBox +\VLin +3\VBit}
  \deflength{\VMerge}    {\VBit}
  \deflength{\HSBox}     {10 mm}
  \deflength{\HSBoxSep}  {1.5 mm}
  \deflength{\HBitSep}   {0.111\HSBox}
  \deflength{\HXor}      {8\HSBox +7\HSBoxSep}
  \deflength{\HDiff}     {4\HSBox +3\HSBoxSep}
  \deflength{\HBlock}    {\HSBox +\HSBoxSep}
  \deflength{\HFigSep}   {15 mm}
  \deflength{\Radius}    {1 mm}
  \deflength{\Tension}   {0.3\VLin}
  \begin{tikzpicture}
  \useasboundingbox (0,8mm) rectangle (\HXor,-\VDiffPos -\VDiff -\VBit -2mm);
  % .................................................................................................
  % Coordonnées
  %
  \coordinate (Xor) at (0,-\VXorPos);
  \coordinate (Diff-0) at (0,-\VDiffPos);
  \coordinate (Diff-1) at (\HDiff +\HSBoxSep,-\VDiffPos);
  \foreach \X in {0,...,7}{
    \coordinate (SBox-\X) at (\X*\HBlock,-\VSBoxPos);
    \coordinate (NameSBox-\X) at ($(SBox-\X)+0.5*(\HSBox,-\VSBox)$);
    \coordinate (Bundle-\X) at (\X*\HBlock +.5\HSBox,0);
    \coordinate (BitA-\X) at (\X*\HBlock +.5\HSBox,0);
    \coordinate (BitB-\X) at (\X*\HBlock +.5\HSBox,-\VXorPos -\VXor);
    \coordinate (BitC-\X) at (\X*\HBlock +.5\HSBox,-\VSBoxPos -\VSBox);
    \coordinate (BitD-\X) at (\X*\HBlock +.5\HSBox,-\VLinPos -\VLin);
    \coordinate (BitE-\X) at (\X*\HBlock +.5\HSBox,-\VDiffPos -\VDiff);
  }
  % .................................................................................................
  % Lignes et boites
  %
  \draw[BoxN] (Xor) rectangle ++(\HXor,-\VXor);
  \draw[BoxN] (Diff-0) rectangle ++(\HDiff,-\VDiff);
  \draw[BoxN] (Diff-1) rectangle ++(\HDiff,-\VDiff);
  \foreach \X in {0,...,7}{
    \draw[BoxN] (SBox-\X) rectangle ++(\HSBox,-\VSBox);
  }
  \foreach \X/\Y in {0/0,1/5,2/2,3/7,4/4,5/1,6/6,7/3}{
    \draw[Merge1] (BitA-\X) -- ++(0,-\VBit);
    \draw[Merge1] (BitB-\X) -- ++(0,-\VBit);
    \draw[Merge1] (BitC-\X) -- ++(0,-\VBit) \BendTo (BitD-\Y) -- ++(0,-\VBit);
    \draw[Merge1] (BitE-\X) -- ++(0,-\VBit);
  }
  \foreach \X in {0,...,7}{
    \draw[LineN,rounded corners=\Radius] (\X*\HBlock,\VMerge)
    -- ++(0,-\VMerge) -- ++(\HSBox, 0) -- ++(0,\VMerge);
    \foreach \Y in {1,...,8}{
      \draw[LineN] (\X*\HBlock +\Y*\HBitSep,0) -- ++(0,\VMerge);
    }
  }
  % .................................................................................................
  % Textes
  %
  \foreach \X/\Txt in {0/0,1/1,2/2,3/3,4/0,5/1,6/2,7/3}{
    \draw[NameN] (NameSBox-\X) node {$S_{\Txt}$};
  }
  \draw[NameN] ($(Diff-0)+0.5*(\HDiff,-\VDiff)$) node {$M$};
  \draw[NameN] ($(Diff-1)+0.5*(\HDiff,-\VDiff)$) node {$M$};
  \draw[NameN] ($(Xor)+0.5*(\HXor,-\VXor)$) node {$\oplus$};
  \draw (0,0) node[above=2mm] {\tiny\makebox[0pt][r]{Bit~}};
  \draw (0,0) node[above=5mm] {\tiny\makebox[0pt][r]{Bundle~}};
  \foreach \X in {0,...,7}{
    \draw (\X*\HBlock +.5\HSBox,0) node[above=2mm] {\tiny\X 0--\X 9};
    \draw (\X*\HBlock +.5\HSBox,0) node[above=5mm] {\tiny\X};
  }
  \end{tikzpicture}
  \caption{A diagrammatic representation of the round function}\label{Fig:Round}%
\end{figure*}
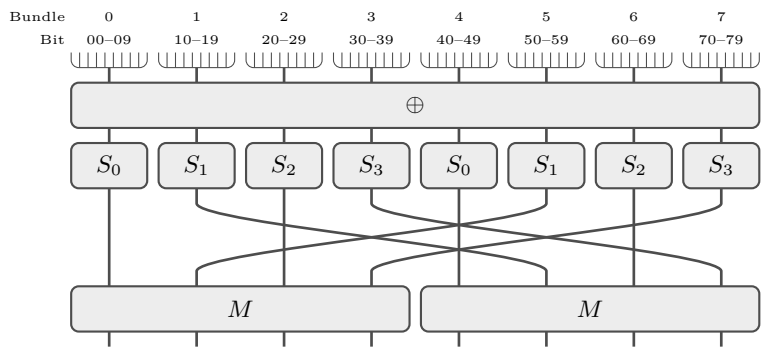

The round function in Figure~\ref{Fig:Round} is made up of three distinct stages: a \emph{key addition}, a \emph{substitution layer} and a \emph{diffusion layer}. The key addition is just a bitwise XOR between the data block and the round key. The substitution layer consists in the parallel evaluation of four different S-boxes and is the only part of the cipher which is not linear or affine. Following the design principles of the AES, the diffusion layer comes in two parts: the \texttt{ShiftRows} and the \texttt{MixColumns} operations.

The decryption is straightforward from the encryption since all the components are bijective. Thus, to decrypt, we just have to apply the inverse operations in the reverse order. Remark that the key addition and the \texttt{ShiftRows} are involutions, therefore the same operations are used in the decryption process.
%------------------------------------------
%\subsection{Description of the Algorithm}
In contrast to the AES, the algorithm works with bundles of 10 bits instead of 8 bits. Let $\FF_2$ denote the Galois Field of order 2. Any $10n$-bit block $x$ is seen as $n$-tuple of 10-bit bundles $(x_0,\ldots,x_{n-1})$, and thus as an element of $(\FF_2^{10})^n$. The hexadecimal notation is used to denote any 10-bit bundle. For example, \hex{37A} stands for 1101111010 in $\FF_2^{10}$.

The S-boxes $S_0$, $S_1$, $S_2$ and $S_3$ are four permutations of $\FF_2^{10}$ given in Appendix. The linear map $M : (\FF_2^{10})^4 \to (\FF_2^{10})^4$ processes four 10-bit bundles. Because of the linearity of this map, $M$ is only defined on the standard basis of $(\FF_2^{10})^4$. For convenience, its inverse $M^{-1}$ is also given in Appendix.

A pseudo-code for the key schedule is given in Algorithm \ref{Algo:ExpKey}. To provide an overview of its structure, the first step is represented in Figure~\ref{Fig:ExpKey}. This representation also emphasizes the similarities between our key schedule and the AES one.
The pseudo-code for the encryption and decryption functions are respectively given in Algorithms \ref{Algo:Encrypt} and \ref{Algo:Decrypt}. The notation $[a \parallel b]$ denotes the concatenation of the vectors $a$ and $b$. Again, an overview of the round function is given in Figure~\ref{Fig:Round}.

\begin{figure*}
  \begin{Algorithm}[Title={ExpandKey}]{\label{Algo:ExpKey}%
      \Input{The 120-bit master key $K = (K_0,\ldots,K_{11}) \in (\FF_2^{10})^{12}$.}
      \Output{The twelve 80-bit round keys $k^0, \ldots, k^{11} \in (\FF_2^{10})^{8}$.}
    }%
    \State{$(k_0,\ldots,k_{11}) \gets (K_0,\ldots,K_{11})$}
    \medskip
    \For{i}{0}{6}
    \State{$x \gets M(k_{12i + 8},\ldots,k_{12i + 11})$}
    \State{$x \gets (S_j(x_j))_{0\leq j \leq 3}$}
    \State{$x \gets (x_0 \oplus (3^i \mod 2^{10}), x_1, x_2, x_3)$}
    \State{$(k_{12i + 12},\ldots,k_{12i + 15})
      \gets (k_{12i + 0},\ldots,k_{12i + 3\hphantom{1}}) \oplus x$}
    \State{$(k_{12i + 16},\ldots,k_{12i + 19})
      \gets (k_{12i + 4},\ldots,k_{12i + 7\hphantom{1}}) \oplus (k_{12i + 12},\ldots,k_{12i + 15})$}
    \State{$(k_{12i + 20},\ldots,k_{12i + 23})
      \gets (k_{12i + 8},\ldots,k_{12i + 11}) \oplus (k_{12i + 16},\ldots,k_{12i + 19})$}
    \EndFor
    \medskip
    \For{r}{0}{11}
    \State{$k^r \gets (k_{8r + i})_{0 \leq i \leq 7}$}
    \EndFor
  \end{Algorithm}
\end{figure*}

\begin{figure*}
  \begin{Algorithm}[Title={Encrypt}]{\label{Algo:Encrypt}%
      \Input{The 120-bit master key $K \in (\FF_2^{10})^{12}$ and the 80-bit plaintext block $p \in (\FF_2^{10})^{8}$.}
      \Output{The 80-bit ciphertext block $c \in (\FF_2^{10})^{8}$.}
    }%
    \State{$k^0,\ldots,k^{11} \gets{}$ExpandKey($K$)}
    \State{$x \gets p$}
    \smallskip
    \For{r}{0}{9}
    \State{$x \gets x \oplus k^r$\strut}
    \Comment{AddRoundKey}
    \State{$x \gets (S_{i \bmod{4}}(x_i))_{0 \leq i \leq 7}$}
    \Comment{SubBundles}
    \State{$x \gets (x_0,x_5,x_2,x_7,x_4,x_1,x_6,x_3)$}
    \Comment{ShiftRows}
    \State{$x \gets [ M(x_0,x_1,x_2,x_3) \parallel M(x_4,x_5,x_6,x_7) ]$}
    \Comment{MixColumns}
    \EndFor
    \smallskip
    \State{$x \gets x \oplus k^{10}$\strut}
    \Comment{AddRoundKey}
    \State{$x \gets (S_{i \bmod{4}}(x_i))_{0 \leq i \leq 7}$}
    \Comment{SubBundles}
    \State{$x \gets (x_0,x_5,x_2,x_7,x_4,x_1,x_6,x_3)$}
    \Comment{ShiftRows}
    \State{$x \gets x \oplus k^{11}$}
    \Comment{AddRoundKey}
    \smallskip
    \State{$c \gets x$\strut}
  \end{Algorithm}
\end{figure*}

\begin{figure*}
  \begin{Algorithm}[Title={Decrypt}]{\label{Algo:Decrypt}%
      \Input{The 120-bit master key $K \in (\FF_2^{10})^{12}$ and the 80-bit ciphertext block $c \in (\FF_2^{10})^{8}$.}
      \Output{The 80-bit plaintext block $p \in (\FF_2^{10})^{8}$.}
    }%
    \State{$k^0,\ldots,k^{11} \gets{}$ExpandKey($K$)}
    \State{$x \gets c$}
    \smallskip
    \State{$x \gets x \oplus k^{11}$\strut}
    \Comment{AddRoundKey}
    \State{$x \gets (x_0,x_5,x_2,x_7,x_4,x_1,x_6,x_3)$}
    \Comment{InvShiftRows}
    \State{$x \gets (S^{-1}_{i \bmod{4}}(x_i))_{0 \leq i \leq 7}$}
    \Comment{InvSubBundles}
    \State{$x \gets x \oplus k^{10}$}
    \Comment{AddRoundKey}
    \smallskip
    \For{r}{9}{0}
    \State{$x \gets [ M^{-1}(x_0,x_1,x_2,x_3) \parallel M^{-1}(x_4,x_5,x_6,x_7) ]$}
    \Comment{InvMixColumns}
    \State{$x \gets (x_0,x_5,x_2,x_7,x_4,x_1,x_6,x_3)$}
    \Comment{InvShiftRows}
    \State{$x \gets (S^{-1}_{i \bmod{4}}(x_i))_{0 \leq i \leq 7}$}
    \Comment{InvSubBundles}
    \State{$x \gets x \oplus k^r$\strut}
    \Comment{AddRoundKey}
    \EndFor
    \smallskip
    \State{$p \gets x$}
  \end{Algorithm}
\end{figure*}

%---------------------------------------------------------

\section{Cryptographic Security Analysis of BEA-1} \label{s3}
\subsection{Differential and Linear Cryptanalyses}
In \cite{daemen2002design}, Daemen and Rijmen introduced the differential and the linear branch numbers of a linear transformation. With an exhaustive search, it can be checked that the differential and linear branch numbers of $M$ are both equal to 5, which is the maximum. This implies that any 2-round trail has at least 5 active S-boxes. Thus, a 10-round trail involves at least 25 active S-boxes. 

Note that all the S-boxes are (at most) differentially 40-uniform and linearly 128-uniform. Therefore, the probability of any 10-round differential trail is upper bounded by $(\frac{40}{1024})^{25} \approx 2^{-116.9}$ and the absolute bias of a 10-round linear trail is upper bounded by $(\frac{128}{512})^{25} = 2^{-50}$. Consequently, a differential cryptanalysis of the 10-round version of our cipher would require at least $2^{117}$ chosen plaintext/ciphertext pairs and a linear cryptanalysis would require $2^{100}$ known plaintext/ciphertext pairs.

Even if this is a rough approximation since it does not take into account the inter-column diffusion provided by the \texttt{ShiftRows} operation, it suffices to prove the cipher practical resistance against classical differential and linear cryptanalyses. In fact, there is only $2^{80}$ different plaintext/ciphertext pairs for a fixed master key.
%--------------------------------------------------
\subsection{Statistical Analysis of BEA-1}
Any cryptographic algorithm must behave as a random generator or at least must exhibit enough randomness properties. Therefore, its outputs for different classes of inputs must pass all the reference statistical testings. The most widely used is the NIST's Statistical Test Suite (STS) \cite{Rukhin01astatistical}.

We have performed the statistical analysis for BEA-1 with respect to all the tests which are implemented in STS. Our encryption has passed all the tests sucessfully. This result is of rather high importance since
\begin{itemize}
  \item STS is the tool recommended by the US government to evaluate statistical properties of any secure encryption algorithm. It is explicitely mandatory to consider it in the industry. 
  \item The presence of our backdoor remains statistically undetectable which proves that if statistical properties are a necessary condition for cryptographic security it is absolutely not a sufficient property. It may be bypassed by considering statistical simulation techniques \cite{DBLP:journals/virology/FiliolJ07}. Algebraic or combinatorial weaknesses moreover remains out of reach from statistical analysis. 
\end{itemize}
\subsection{Cryptographic Challenge}
We propose a cryptographic challenge whose aims is twofold:
\begin{itemize}
\item identifying and explaing what our backdoor consists in,
\item exploiting this backdoor in the most efficient way (in terms of computing time, memory requirements, the number of required plaintext/ciphertext pairs). 
\end{itemize}
We have run our own full cryptanalysis implementation several times. Each time, we retrieve the 120-bit key successfully. 

This challenge will be officially launched right after the presentation of the present paper, on the \url{arxiv.org}. To take part, participants must send the following data to both authors (prior to any publication):
\begin{itemize}
  \item the description of the backdoor,
  \item the description of the attack to exploit the backdoor successfully. 
  \item the relevant source codes. They will help us to sort the different proposals with respect to, first the number of required plaintext/ciphertext pairs, second the computing time on our reference computer.
\end{itemize}
Incentive (non monetary) awards will be awarded to the three best attacks. Our attack as the reference solution will be presented in at the RusCrypto 2017 conference around end of March 2017. Consequently, the challenge holds until this date. Moreover the best attack will be considered for publication in the Journal in Computer Virology and Hacking Techniques.

%---------------------------------------------------------
\section{Conclusion and Future Work} \label{s4}
In this paper, we have proposed an AES-like encryption algorithm which contains a backdoor at its design level. This algorithm, named BEA-1, exhibits many of the desirable properties that any secure algorithm should. However, it is absolutely unsuitable for actually protection information. Indeed, we manage to break it with a rather limited amount of resources successfully.

While it is a humble, first step in a larger research work, it illustrates the issue of using foreign encryption algorithms which may contains such hidden weaknesses. The very final aim of our work is to prove that it is feasible to embed such undetectable intended weaknesses. It is consequently a critical issue to have a broader work conducted in this  research area and we hope that other people will also consider it as such. 

The next step will be to consider more sophisticated combinatorial structures.

\bibliography{Forse2017}

%------------------------------------------------------------------------------------------------------------------------

\appendix

\section*{Appendix}

\noindent The present appendix contains the different tables for the S-boxes (Figures~\ref{sb0-1} and \ref{sb2-3}), the linear map $M$ and its inverse $M^{-1}$ (Figure~\ref{table_m}). They can be copied and pasted for a practical implementation of the encryption algorithm.

\begin{figure}[!b]
\newcommand{\NULL}{\hex{\color{black!30}000}}
\centering
\begin{footnotesize}
  \begin{math}%
\begin{array}{ @{} c @{\;\mapsto\;} c@{} }
  x & M(x) \\\hline\rule{0pt}{3ex}%
  (\hex{001}, \NULL, \NULL, \NULL) & (\hex{112}, \hex{1BC}, \hex{36C}, \hex{0C5}) \\
  (\hex{002}, \NULL, \NULL, \NULL) & (\hex{344}, \hex{394}, \hex{342}, \hex{165}) \\
  (\hex{004}, \NULL, \NULL, \NULL) & (\hex{23F}, \hex{15B}, \hex{0C7}, \hex{0A7}) \\
  (\hex{008}, \NULL, \NULL, \NULL) & (\hex{215}, \hex{11F}, \hex{1E0}, \hex{2E7}) \\
  (\hex{010}, \NULL, \NULL, \NULL) & (\hex{2D9}, \hex{10A}, \hex{0C4}, \hex{095}) \\
  (\hex{020}, \NULL, \NULL, \NULL) & (\hex{231}, \hex{120}, \hex{322}, \hex{016}) \\
  (\hex{040}, \NULL, \NULL, \NULL) & (\hex{3C6}, \hex{010}, \hex{0EC}, \hex{261}) \\
\end{array}%
  \end{math}%
  \hfil
  \begin{math}%
  \begin{array}{ @{} c @{\;\mapsto\;} c@{} }
  x & M^{-1}(x) \\\hline\rule{0pt}{3ex}%
  (\hex{001}, \NULL, \NULL, \NULL) & (\hex{10B}, \hex{221}, \hex{09D}, \hex{398}) \\
  (\hex{002}, \NULL, \NULL, \NULL) & (\hex{1AE}, \hex{1E9}, \hex{2CB}, \hex{245}) \\
  (\hex{004}, \NULL, \NULL, \NULL) & (\hex{1AB}, \hex{11E}, \hex{05F}, \hex{3A4}) \\
  (\hex{008}, \NULL, \NULL, \NULL) & (\hex{08D}, \hex{04D}, \hex{016}, \hex{34C}) \\
  (\hex{010}, \NULL, \NULL, \NULL) & (\hex{0AD}, \hex{337}, \hex{3C5}, \hex{2D4}) \\
  (\hex{020}, \NULL, \NULL, \NULL) & (\hex{322}, \hex{3FD}, \hex{3D5}, \hex{0E5}) \\
  (\hex{040}, \NULL, \NULL, \NULL) & (\hex{002}, \hex{246}, \hex{2E2}, \hex{380}) \\
\end{array}%
  \end{math}%
\end{footnotesize}
\caption{Specification of $M$ and $M^{-1}$} \label{table_m}
\end{figure}

\newlength{\sep}\deflength{\sep}{1.3mm}
\renewcommand{\arraystretch}{1.25}
\begin{figure}[!ht]
\centering
\begin{tiny}\begin{tabular}{ @{} c @{\hspace{.9\sep}} | @{\hspace{.9\sep}} *{15}{c @{\hspace{\sep}} } c @{} }
  $S_0$     & \hex{..0} & \hex{..1} & \hex{..2} & \hex{..3} & \hex{..4} & \hex{..5} & \hex{..6} & \hex{..7} &
  \hex{..8} & \hex{..9} & \hex{..A} & \hex{..B} & \hex{..C} & \hex{..D} & \hex{..E} & \hex{..F} \\
  \hline\rule{0pt}{3ex}%
  \hex{00.} & \hex{0BA} & \hex{026} & \hex{0A0} & \hex{1E1} & \hex{183} & \hex{3DB} & \hex{1A4} & \hex{083} &
  \hex{110} & \hex{350} & \hex{085} & \hex{2E5} & \hex{3B4} & \hex{195} & \hex{359} & \hex{2E6} \\
  \hex{01.} & \hex{33A} & \hex{26B} & \hex{209} & \hex{217} & \hex{1CE} & \hex{2E3} & \hex{0C0} & \hex{136} &
  \hex{129} & \hex{0C8} & \hex{3D6} & \hex{054} & \hex{040} & \hex{3F2} & \hex{09F} & \hex{322} \\
  \hex{02.} & \hex{11B} & \hex{07F} & \hex{139} & \hex{07D} & \hex{2CF} & \hex{02A} & \hex{268} & \hex{227} &
  \hex{246} & \hex{1C5} & \hex{12B} & \hex{3B6} & \hex{16C} & \hex{20D} & \hex{1E7} & \hex{35B} \\
  \hex{03.} & \hex{313} & \hex{0CD} & \hex{11E} & \hex{1E6} & \hex{117} & \hex{355} & \hex{182} & \hex{0E6} &
  \hex{094} & \hex{1B9} & \hex{19C} & \hex{28C} & \hex{2B9} & \hex{336} & \hex{0AF} & \hex{19D} \\
  \hex{04.} & \hex{2BC} & \hex{1A9} & \hex{31B} & \hex{02E} & \hex{282} & \hex{2AE} & \hex{272} & \hex{2E9} &
  \hex{3AA} & \hex{1DD} & \hex{013} & \hex{2D3} & \hex{30F} & \hex{35A} & \hex{159} & \hex{1BB} \\
  \hex{05.} & \hex{11C} & \hex{12A} & \hex{248} & \hex{3C7} & \hex{28B} & \hex{191} & \hex{025} & \hex{173} &
  \hex{018} & \hex{38D} & \hex{1A1} & \hex{185} & \hex{007} & \hex{156} & \hex{378} & \hex{312} \\
  \hex{06.} & \hex{0C9} & \hex{143} & \hex{05D} & \hex{3FA} & \hex{038} & \hex{3DE} & \hex{081} & \hex{0F9} &
  \hex{2D1} & \hex{3FB} & \hex{1C7} & \hex{3E0} & \hex{1DC} & \hex{16A} & \hex{2D8} & \hex{23F} \\
\end{tabular}
\hfill
\begin{tabular}{ @{} c @{\hspace{.9\sep}} | @{\hspace{.9\sep}} *{15}{c @{\hspace{\sep}} } c @{} }
  $S_1$     & \hex{..0} & \hex{..1} & \hex{..2} & \hex{..3} & \hex{..4} & \hex{..5} & \hex{..6} & \hex{..7} &
  \hex{..8} & \hex{..9} & \hex{..A} & \hex{..B} & \hex{..C} & \hex{..D} & \hex{..E} & \hex{..F} \\
  \hline\rule{0pt}{3ex}%
  \hex{00.} & \hex{021} & \hex{09B} & \hex{37A} & \hex{3AB} & \hex{0DF} & \hex{016} & \hex{1FE} & \hex{004} &
  \hex{07C} & \hex{3BE} & \hex{141} & \hex{397} & \hex{300} & \hex{185} & \hex{00C} & \hex{1A7} \\
  \hex{01.} & \hex{2FA} & \hex{3AA} & \hex{235} & \hex{0B9} & \hex{003} & \hex{3CF} & \hex{14A} & \hex{18F} &
  \hex{356} & \hex{363} & \hex{173} & \hex{2E4} & \hex{168} & \hex{0CF} & \hex{373} & \hex{379} \\
  \hex{02.} & \hex{2CA} & \hex{326} & \hex{16B} & \hex{393} & \hex{283} & \hex{2E0} & \hex{2B9} & \hex{3E9} &
  \hex{12F} & \hex{247} & \hex{3D8} & \hex{07B} & \hex{288} & \hex{146} & \hex{30F} & \hex{267} \\
  \hex{03.} & \hex{15C} & \hex{01F} & \hex{22C} & \hex{0F8} & \hex{10F} & \hex{35D} & \hex{367} & \hex{343} &
  \hex{1EC} & \hex{047} & \hex{008} & \hex{062} & \hex{2CF} & \hex{3D6} & \hex{36B} & \hex{148} \\
  \hex{04.} & \hex{0B4} & \hex{2E3} & \hex{25E} & \hex{234} & \hex{0D2} & \hex{1F8} & \hex{184} & \hex{2FF} &
  \hex{2EB} & \hex{2BB} & \hex{3A1} & \hex{34F} & \hex{312} & \hex{10B} & \hex{2EA} & \hex{04D} \\
  \hex{05.} & \hex{1B1} & \hex{2FE} & \hex{084} & \hex{229} & \hex{216} & \hex{337} & \hex{0D4} & \hex{08D} &
  \hex{21F} & \hex{035} & \hex{164} & \hex{32A} & \hex{1AA} & \hex{182} & \hex{24B} & \hex{1BF} \\
  \hex{06.} & \hex{245} & \hex{257} & \hex{01E} & \hex{34E} & \hex{375} & \hex{197} & \hex{292} & \hex{1DD} &
  \hex{14D} & \hex{190} & \hex{27E} & \hex{13D} & \hex{137} & \hex{3A3} & \hex{228} & \hex{392} \\
\end{tabular}
\end{tiny}
\caption{Specification of the S-boxes $S_0$ and $S_1$} \label{sb0-1}
\end{figure}
    
\begin{figure}[!ht]
\centering
\begin{tiny}\begin{tabular}{ @{} c @{\hspace{.9\sep}} | @{\hspace{.9\sep}} *{15}{c @{\hspace{\sep}} } c @{} }
  $S_2$     & \hex{..0} & \hex{..1} & \hex{..2} & \hex{..3} & \hex{..4} & \hex{..5} & \hex{..6} & \hex{..7} &
  \hex{..8} & \hex{..9} & \hex{..A} & \hex{..B} & \hex{..C} & \hex{..D} & \hex{..E} & \hex{..F} \\
  \hline\rule{0pt}{3ex}%
  \hex{00.} & \hex{12E} & \hex{38B} & \hex{18E} & \hex{131} & \hex{039} & \hex{10D} & \hex{2DE} & \hex{246} &
  \hex{286} & \hex{2BE} & \hex{315} & \hex{384} & \hex{21D} & \hex{142} & \hex{06D} & \hex{0CA} \\
  \hex{01.} & \hex{2A2} & \hex{2CE} & \hex{264} & \hex{085} & \hex{374} & \hex{3BB} & \hex{3B9} & \hex{1B7} &
  \hex{3E6} & \hex{3BC} & \hex{207} & \hex{002} & \hex{392} & \hex{1B5} & \hex{0BA} & \hex{318} \\
  \hex{02.} & \hex{39C} & \hex{2EE} & \hex{1DA} & \hex{125} & \hex{019} & \hex{063} & \hex{27E} & \hex{126} &
  \hex{19A} & \hex{082} & \hex{305} & \hex{0E3} & \hex{206} & \hex{0A0} & \hex{009} & \hex{3C6} \\
  \hex{03.} & \hex{100} & \hex{3F3} & \hex{2AD} & \hex{199} & \hex{102} & \hex{108} & \hex{1DB} & \hex{2F0} &
  \hex{310} & \hex{245} & \hex{0A8} & \hex{116} & \hex{022} & \hex{3C1} & \hex{028} & \hex{332} \\
  \hex{04.} & \hex{1E1} & \hex{2E7} & \hex{0DA} & \hex{2B7} & \hex{0CB} & \hex{07C} & \hex{2A0} & \hex{240} &
  \hex{150} & \hex{165} & \hex{258} & \hex{2C8} & \hex{0C4} & \hex{334} & \hex{36B} & \hex{2D1} \\
  \hex{05.} & \hex{1E0} & \hex{138} & \hex{39A} & \hex{0FF} & \hex{1A7} & \hex{10C} & \hex{353} & \hex{19B} &
  \hex{171} & \hex{038} & \hex{3BD} & \hex{000} & \hex{3A2} & \hex{1B8} & \hex{282} & \hex{2EB} \\
  \hex{06.} & \hex{1D4} & \hex{3D4} & \hex{20F} & \hex{23C} & \hex{0D7} & \hex{154} & \hex{012} & \hex{0DF} &
  \hex{3A8} & \hex{237} & \hex{09E} & \hex{155} & \hex{2E2} & \hex{189} & \hex{2F2} & \hex{136} \\
\end{tabular}
\hfill
\begin{tabular}{ @{} c @{\hspace{.9\sep}} | @{\hspace{.9\sep}} *{15}{c @{\hspace{\sep}} } c @{} }
  $S_3$     & \hex{..0} & \hex{..1} & \hex{..2} & \hex{..3} & \hex{..4} & \hex{..5} & \hex{..6} & \hex{..7} &
  \hex{..8} & \hex{..9} & \hex{..A} & \hex{..B} & \hex{..C} & \hex{..D} & \hex{..E} & \hex{..F} \\
  \hline\rule{0pt}{3ex}%
  \hex{00.} & \hex{200} & \hex{084} & \hex{1B5} & \hex{30A} & \hex{25A} & \hex{151} & \hex{174} & \hex{3F9} &
  \hex{113} & \hex{3B4} & \hex{35B} & \hex{291} & \hex{332} & \hex{170} & \hex{021} & \hex{31E} \\
  \hex{01.} & \hex{00E} & \hex{2FC} & \hex{023} & \hex{0B0} & \hex{3A9} & \hex{259} & \hex{2BC} & \hex{378} &
  \hex{031} & \hex{050} & \hex{0D0} & \hex{1FF} & \hex{26C} & \hex{0D5} & \hex{214} & \hex{23E} \\
  \hex{02.} & \hex{1AB} & \hex{0AB} & \hex{3AC} & \hex{036} & \hex{0E2} & \hex{2F6} & \hex{07A} & \hex{0EA} &
  \hex{2CB} & \hex{0FE} & \hex{24E} & \hex{280} & \hex{138} & \hex{073} & \hex{219} & \hex{3EA} \\
  \hex{03.} & \hex{2E2} & \hex{27C} & \hex{032} & \hex{162} & \hex{285} & \hex{13C} & \hex{0B6} & \hex{1ED} &
  \hex{0B3} & \hex{2F5} & \hex{2C6} & \hex{34B} & \hex{335} & \hex{1EF} & \hex{26E} & \hex{37A} \\
  \hex{04.} & \hex{273} & \hex{17E} & \hex{30F} & \hex{2E7} & \hex{14B} & \hex{3BC} & \hex{1CE} & \hex{039} &
  \hex{315} & \hex{01A} & \hex{144} & \hex{1C4} & \hex{20A} & \hex{17B} & \hex{362} & \hex{10D} \\
  \hex{05.} & \hex{235} & \hex{1D9} & \hex{2F9} & \hex{0A4} & \hex{052} & \hex{0E3} & \hex{0BD} & \hex{061} &
  \hex{02C} & \hex{140} & \hex{0E1} & \hex{156} & \hex{10E} & \hex{250} & \hex{288} & \hex{1BE} \\
  \hex{06.} & \hex{07C} & \hex{2B8} & \hex{05D} & \hex{242} & \hex{192} & \hex{0A8} & \hex{3B0} & \hex{0DB} &
  \hex{129} & \hex{2AF} & \hex{063} & \hex{3AF} & \hex{3D1} & \hex{0C8} & \hex{0A6} & \hex{029} \\
\end{tabular}\end{tiny}
\caption{Specification of the S-boxes $S_2$ and $S_3$} \label{sb2-3}
\end{figure}
\end{document}